\documentclass[12pt]{article}
\usepackage{epsfig}
\usepackage{GGGraphics}
\usepackage{graphicx}
\usepackage{amsfonts}

\def\functionspace{\mathfrak}

\def\P{\mathfrak P}
\def\O{\mathcal O}
\def\X{\mathcal X}
\def\Y{\mathcal Y}
\def\Z{\mathcal Z}

\def\R{{\mathbb R}}
\def\C{{\mathbb C}}

\def\X{{\functionspace X}}

\def\div{{\rm div}}
\def\dim{\mathop{\rm dim}}

\newtheorem{definition}{Definition}
\newtheorem{conjecture}{Conjecture}

\newtheorem{theorem}{Theorem}

\newcommand{\QED}{\hspace*{2em}\hfill$\bullet$}
\newcommand{\order}{\mathop{\rm order}}
\newcommand{\degree}{\mathop{\rm degree}}
\newenvironment{proof}[1]{\vspace{-2pt}{\em Proof\/}\quad #1}{\QED
  \vspace{8pt}}

\def\eqalign#1{\null\,\vcenter{\openup\jot \mathsurround=0pt
  \ialign{\strut\hfil$\displaystyle{##}$&$
        \displaystyle{{}##}$\hfil \crcr#1\crcr}}\,}

\title{The algebraic entropy of classical mechanics}
\author{Robert I. McLachlan and Brett Ryland\thanks{Institute of 
Fundamental Sciences, 
Massey University, Palmerston North, New Zealand 
({\tt R.McLachlan@massey.ac.nz}).}
\date{\normalsize \it Dedicated to Gerhard Wanner on the occasion of his 60th
birthday. \\ Of trees and the counting of trees, may there be no end!}
}
\begin{document}

\maketitle

\begin{abstract}
\noindent
We describe the `Lie algebra of classical mechanics', modelled on the
Lie algebra generated by kinetic and potential energy of a simple
mechanical system  with respect to the canonical Poisson bracket.
It is a polynomially graded Lie algebra, a class we introduce.
We describe these Lie algebras, 
give an algorithm to calculate the dimensions $c_n$ of the homogeneous
subspaces of the Lie algebra of classical mechanics,
and determine the value of its entropy $\lim_{n\to\infty} c_n^{1/n}$.
It is  $1.82542377420108\dots$, a fundamental constant associated to
classical mechanics.
\end{abstract}

%


\def\eqalign#1{\null\,\vcenter{\openup\jot \mathsurround=0pt
  \ialign{\strut\hfil$\displaystyle{##}$&$
        \displaystyle{{}##}$\hfil \crcr#1\crcr}}\,}

\def\epsilon{\varepsilon}
\def\phi{\varphi} \def\A{{\cal{A}}}
\def\fix{\mathop{\rm fix}}

\def\smallmatrix{\null\,\vcenter\bgroup \baselineskip=7pt
 \ialign\bgroup\hfil$\scriptstyle{##}$\hfil&&$\;$\hfil
 $\scriptstyle{##}$\hfil\crcr}
\def\endsmallmatrix{\crcr\egroup\egroup\,}
\def\smatrix#1{\smallmatrix #1 \endsmallmatrix}
\def\spmatrix#1{\left( \smallmatrix #1 \endsmallmatrix \right)}

\newpage
\baselineskip=18pt
\section{Introduction. Classes of Lie algebras.}
The class of `simple mechanical systems' are defined by pairs $(Q,V)$, where
the configuration space $Q$ is a real Riemannian manifold and the 
potential energy $V$ is a smooth real function on $Q$.
The phase space $T^*Q$ has a canonical
Poisson bracket and a kinetic energy $T:T^*Q\to \R$ associated with the
metric on $Q$. In general, the smooth functions on a Poisson manifold
form a Lie algebra under the Poisson bracket. In the case of
a simple mechanical system, we are given two distinguished functions,
namely the kinetic and potential energies, and one can ask what 
Lie algebra they generate under the Poisson bracket. 

In this paper we study, not the Lie algebra generated by a {\it particular}
$V$ and $T$, but the Lie algebra defined by the whole {\it class}
of simple mechanical systems. That is, one should think of the dimension
of $Q$ as being arbitrarily large, and the metric and potential energies
also being arbitrary. 

This question arose out of very practical considerations of the calculations
required to derive high-order symplectic integrators by splitting and
composition, used in applications including molecular, celestial, 
and accelerator dynamics \cite{mc-qu,ha-lu-wa}. 
The vector field $X$ which is to be integrated
is split as $X=A+B$, where $A$ and $B$ have the same properties (e.g.
Hamiltonian) as $X$, but can be integrated exactly.
We write $\exp(tX)$ for the time-$t$ flow of $X$.
The most common such integrator is the leap-frog method
$$\phi(\tau) := \exp(\frac{1}{2}\tau A)\exp(\tau B)\exp(\frac{1}{2}\tau A),$$
where the small parameter $\tau$ is the time step.

From the Baker-Campbell-Hausdorff formula \cite{bourbaki}, the map $\phi(\tau)$
can be represented (up to any power in $\tau$) as a flow 
$\exp(\tau \widetilde X)$, where
\begin{equation}
\label{eq:aba}
\widetilde 
X = A+B+\tau^2(\frac{1}{12}[B,[B,A]] -\frac{1}{24}[A,[A,B]]))
+ \O(\tau^4).
\end{equation}
Because it is the flow of
a vector field $\O(\tau^2)$-close to the original one, the integrator
is second order. The function
$\widetilde X$ is called the {\it modified vector field} in the
numerical integration literature \cite{ha-lu-wa}.

For simple mechanical systems, we split the Hamiltonian as $H=T+V$.
The flow of (the Hamiltonian vector field of) $V$ can of course 
always be calculated easily, but calculating the flow of 
the kinetic energy $T$ requires 
that $Q$ have integrable (and even fairly simple) geodesics.
Because the Lie algebras of Hamiltonian vector fields and
of Hamiltonian functions are isomorphic under $[X_T,X_V]=X_{\{V,T\}}$,
there is a series formally identical to Eq. (\ref{eq:aba}) involving the
Hamiltonians $T$ and $V$ with respect to the Poisson bracket.

In the series of Eq. (\ref{eq:aba}) we see the Lie algebra generated by $A$ and $B$
entering. Such series, for example in the proof of the BCH formula, are usually
considered in the context of the free Lie algebra $L(A,B)$ 
with two generators $A$ and $B$.
One can in fact consider the more general composition
\begin{equation}
\label{eq:higher}
\prod_{i=1}^s \exp(a_i \tau A)\exp(b_i \tau B) = \exp(Z)
\end{equation}
where $Z\in L(A,B)$. Requiring $Z=\tau(A+B)+O(\tau^{p+1})$ for
some integer $p>1$ gives a system of equations in the $a_i$ and $b_i$
which must be satisfied for the method to have order $p$. In the
case of general $A$ and $B$, then, at each order $n=1,\dots,p$ there are
$\dim L_n(A,B)$ such {\it order conditions}. Here $L_n(A,B)$ is the
subspace of $L(A,B)$ consisting of homogeneous elements of order $n$.
Witt's formula \cite{bourbaki} states that
\begin{equation}
\label{eq:witt}
\dim L_n(A,B) = \frac{1}{n}\sum_{d|n}\mu(d) 2^{n/d}
\end{equation}
where $\mu(d)$ is the M\"obius function defined by
$\mu(1) = 1$, $\mu(d) = (-1)^k$ if $d$ is the product of
$k$ distinct primes, and $\mu(d)=0$ otherwise.
Notice that in this case
$$\dim L_n(A,B) \sim \frac{2^n}{n};$$
the dimensions grow exponentially with $n$. The base (2 in this case) 
of the exponent
is called the {\it entropy} of $L(A,B)$. In general, the entropy of
a graded vector space $\bigoplus L_n$ is 
$$\limsup_{n\to\infty}\,(\dim L_n)^{1/n},$$
if this limit exists \cite{newman}. (We shall use generalizations of 
Witt's formula
\cite{ka-ki,mu-ow} to calculate the dimensions and
entropies of other free Lie algebras, see Eqs. 
(\ref{eq:cns}), (\ref{eq:cns2}) below.)

In this approach it is assumed that there are no Lie identities 
satisfied by the vector fields 
$A$ and $B$. This is reasonable if one wants the method to work
for all $A$ and $B$. However, in the case of simple mechanical systems,
the Lie algebra is {\it never} free,
regardless of $T$, $V$, or the dimension
of the system. There are always identities satisfied by kinetic and potential
energy. The simplest
of these is 
\begin{equation}
\label{eq:vvvt}
\{V,\{V,\{V,T\}\}\}\equiv 0.
\end{equation} 
For, working in local
coordinates $(q,p)$ with $T=\frac{1}{2}p^T M(q) p$, and recalling
the canonical Poisson bracket
$\{A,B\} := \sum_i
\frac{\partial A}{\partial q_i}\frac{\partial B}{\partial p_i} -
\frac{\partial A}{\partial p_i}\frac{\partial B}{\partial q_i},$
we have that
$$ \{V,T\} = \sum_{i,j} \frac{\partial V}{\partial q_i} M_{ij}(q) p_j $$
is of degree 1 in $p$, and that
\begin{equation}
\label{eq:vvt}
\{V,\{V,T\}\} = \sum_{i,j}\frac{\partial V}{\partial q_i} M_{ij} 
\frac{\partial V}{\partial q_j}
\end{equation}
is a function of $q$ only. So $V$ and $\{V,\{V,T\}\}$ commute.

Thus, it was realized early on \cite{mclachlan} that in deriving 
high-order integrators as in Eq. (\ref{eq:higher}) for simple
mechanical systems, the order conditions
corresponding to $\{V,\{V,\{V,T\}\}\}$ and to all its higher Lie brackets can
be dropped. This means that more efficient integrators can be designed
for this class of systems. Much work has been done on this special case,
both because of its intrinsic theoretical and practical 
importance, and because it allows such big improvements over the 
general case. For example,
one can design special (`corrector' or `processor') 
methods of the form $\phi\psi\phi^{-1}$
\cite{bl-ca-ro1},
special methods for nearly-integrable systems such as the solar system
\cite{bl-ca-ro2,wi-ho-to},
special methods involving exact evaluation of the forces associated
with the `modified potential' (Eq. \ref{eq:vvt}) \cite{bl-ca-ro3}, and so on---see
\cite{mc-qu} for a survey. All of these studies
rely on the structure of the Lie algebra generated by kinetic and potential
energy.
Bases for this Lie algebra have been constructed, more or less by hand,
for small orders \cite{bl-ca-ro3,bl-mo,murua}. 
In particular, Murua \cite{murua} associates a unique tree of a certain
type to each independent order condition of symplectic Runge-Kutta-Nystr\"om
methods (very closely related to the problem considered here), and enumerates
these up to order 6. (Iserles et al. \cite{sofroniou} extend this approach
to some other classes of polynomial vector fields.)
However, a systematic description of the entire
Lie algebra is clearly preferred.

Not many classes of Lie algebras have been completely described.
Here are two examples from the literature.
First, Duchamp and Krob \cite{du-kr}
completely describe all partially-commutative Lie algebras
$$L(A_1,\dots,A_n; [A_i,A_j]=0, (i,j)\in C)$$
where $C$ specifies the pairs of commuting variables.
Second,
Kirillov, Kontsevich, and Molev
 \cite{kirillov} studied the Lie algebra $L$ generated by two
vector fields on $\R$ in general position, conjectured that
\begin{equation}
\label{eq:standard4}
\sum_{\sigma\in S_4}(-1)^{\mathop{\rm sgn}(\sigma)}[x_{\sigma(1)},
x_{\sigma(2)},x_{\sigma(3)}, x_{\sigma(4)},y]=0
\quad\forall x_1,x_2,x_3,x_4,y\in L
\end{equation}
generates all identities, and calculated the dimensions of its homogeneous
subspaces and the asymptotic growth of their dimension.
If their conjecture is true, $L$
is a {\it PI-algebra} \cite{bahturin,drensky}, 
one which the identities which hold in the Lie algebra 
(such as Eq. (\ref{eq:standard4})) 
are satisfied by {\it all} elements of the Lie
algebra. 

Returning to the case of simple mechanical systems, it is
clear that 
every Lie bracket of $T$ and $V$ is a homogeneous polynomial in $p$.
Furthermore,
the degrees of these polynomials 
combine in a natural way. We therefore introduce the following class $\P$
of Lie algebras.

We use the notation $[XY] := [X,Y]$, $[XYZ]:=[X,[Y,Z]]$, and for
sets $\X$, $\Y$, $[\X\Y] :=[\X,\Y] := \{[X,Y]\colon X\in\X,Y\in\Y\}$.

\begin{definition}
A Lie algebra $L$ is of class $\P$ (`polynomially graded') if it is graded, i.e.
$L=\bigoplus_{n\ge 0}L_n$, and its homogeneous subspaces $L_n$ satisfy
\begin{equation}
\label{eq:degree}
\eqalign{&[L_n,L_m]\subseteq L_{n+m-1}\hbox{\rm\ 
if $n>0$ or $m>0$; and} \cr
   &[L_0,L_0] = 0\cr}
\end{equation}
Note that this implies
$[(L_0)^{n+1}L_n]=0$ for all $n$. 
We call the grading of $L$ its {\rm grading by degree}.
\end{definition}

For example, the Lie algebra generated by kinetic and potential energy
is of class $\P$, where the grading is by total degree in $p$. 
The Lie algebra of all polynomial vector fields on a linear space
is of class $\P$, where the grading is by total degree.
We will give more examples later.

Such a grading is quite different from the natural grading of
a free Lie algebra. Two important differences are that
(i) It is not abelian. For,
$[L_2,[L_0,L_0]] = 0$ while
$[L_0,[L_0,L_2]]\subseteq L_0$. (ii) It is not finite, in the
sense that elements of $L_n$ are Lie brackets of unboundedly
many other elements of $L$. For example, the bracket of any number
of elements of degree 1 is still of degree 1.

We also need the concept of a Lie algebra which is free in a certain class.

\begin{definition}\cite{drensky} Let $F$ be a Lie algebra of class $\P$
generated by a set $\X$. Then $F$ is called a free Lie algebra in
the class $\P$, freely generated by the set $\X$, if for any Lie algebra
$R$ of class $\P$, every mapping $\X\to R$ can be extended to 
a unique homomorphism $F \to R$. We write $F=L_\P(\X)$.
\end{definition}

In addition to the grading by degree, $L_\P(\X)$ also carries
the standard grading which we call the grading by {\it order},
generated by $\order(X)=1$ for all generators $X\in\X$ and 
$\order([Y,Z]) = \order(Y)+\order(Z)$.
(The term {\it order} is chosen here because it corresponds to order
in the sense of numerical integrators, as in Eq. (\ref{eq:aba})).

Because of the importance of the grading by degree for Lie algebras
generated by kinetic and potential energy, we make the following definition.

\begin{definition}The Lie algebra $L_\P(A,B)$, free in the class $\P$,
where $A$ has degree 2 and $B$ has degree 0, is called the
{\rm Lie algebra of classical mechanics}.
\end{definition}

Two Lie algebras of class $\P$ are easy to describe. First, the Lie algebra with $k$
generators of degree $\ge 1$ which is free in the class $\P$ is just
the standard free Lie algebra on $k$ generators---the degrees can never
decrease if the Lie algebra has no elements of degree 0. 
Second,
the Lie algebra with generators $\X=\{X_1,\dots,X_k\}$ of degree 0 and generators 
$\Y=\{Y_1,\dots,Y_l\}$ of
degree 1, free in the class $\P$, is 
$\Y\oplus \bigoplus_{n\ge 0}[\Y^n\X]$,
and only contains elements of degree 0 and 1.
(In both of these cases, the grading by degree
is in fact abelian.)

However, we want to describe the Lie algebra of classical mechanics,
$L_{\P}(A,B)$. This is the simplest nontrivial case as it 
includes the essential feature of $\P$ that degrees can both
increase and decrease under Lie brackets.

The paper is organized as follows. In Section 2, we give a construction
which describes $L_\P(A,B)$ as the direct sum of an abelian
and a free Lie algebra, both with an infinite number of generators.
In Section 3, we enumerate the dimensions
of the homogeneous (by order) components of $L_\P(A,B)$ and hence
in Section 4 numerically compute its entropy.
Section 5 considers special cases (e.g., of mechanical systems
with Euclidean metric; these turn out not to be free in the class $\P$) 
and other examples of polynomially-graded
Lie algebras.

\section{Structure of the Lie algebra of classical mechanics.}

Let $\psi:L(A,B)\to L_\P(A,B)$ be the unique homomorphism from
the free Lie algebra to the free Lie algebra of class $\P$.
The kernel $\ker\psi$ can be thought of as the set of identities
of $L_\P(A,B)$. For example, we showed above (Eq. (\ref{eq:vvvt}))
that $[BBBA]\in\ker\psi$. This implies that $[CBBBA]\in\ker\psi$
for all $C\in L(A,B)$. However, we will see below that
$[BBBA]$ is not the only generator of the ideal $\ker\psi$.

Our description of $L_\P(A,B)$ is based on the following two observations.
First, suppose one wants to describe the Lie algebra with three generators
$A,B,C$ which is free in the class of Lie algebras with $C=0$. Since $C$ generates
all identities in this class, this Lie algebra is just 
$L(A,B)$: one merely has to drop the generator $C$.
To generalize this idea, 
suppose the free Lie algebra $L(A,B)$ can be factored
as $\bigoplus_i L(\X_i)$ for certain generating sets $\X_i$ with elements
in $L(A,B)$, such that some subset $\Y$ of $\cup_i \X_i$ generates all the
identities in $\P$. Then, we have 
\begin{equation}
\label{eq:sum}
L_{\P}(A,B) \cong \bigoplus_i L(\X_i \backslash (\Y\cap \X_i))
\end{equation}
---again, we merely drop these generators.

If $\Y$ only generates {\it some} of the identities of $\P$, then dropping these
generators gives a sum of free Lie algebras which is surjectively homomorphic
to $L_{\P}(A,B)$. This can be used to get upper bounds for the dimensions
of the homogeneous subspaces of $L_{\P}(A,B)$.

Second, given a description of $L_{\P}(A,B)$ as such a sum 
(Eq. \ref{eq:sum}) of free
Lie algebras, we can apply standard techniques to describe it in detail,
for example to construct bases, to compute its dimensions with respect 
to degree 
and/or order, and to compute the asymptotic growth of these dimensions.

We begin by stating the crucial tool we shall use, the Lazard factorization of
free Lie algebras.
\begin{theorem}\cite{lazard,bourbaki,lothaire}
Let $\X$ and $\Y$ be sets of generators. Then
$$L(\X\cup\Y) \cong L(\Y) \oplus L(\cup_{n\ge 0}[\Y^n \X]).$$
\end{theorem}

Applying the Lazard factorization to $L(A,B)$ with $\X=\{A\}$,
$\Y=\{B\}$, gives
$$L(A,B) = B \oplus L(A,[BA],[BBA],[BBBA],\dots)$$
where the elements $[B^nA]$ for $n\ge 3$ are all identities
in $\P$. Thus, $L_\P(A,B)$ is surjectively homomorphic 
to $B\oplus L(A,[BA],[BBA])$. The
three generators have degrees 2 ($A$), 1 ($[BA]$), and 0 ($[BBA]$). 
The idea now is to eliminate this new element of degree 0.
(Formally, the generators $[B^nA]$, $n\ge 3$, do remain in the generating
set; but they and all succeeding Lie brackets of them will  be dropped
at the final stage when we pass to $L_{\P}(A,B)$, so we do not need to
keep track of them and just indicate them by $*$.) This gives
$$\eqalign{ L(A,B) & \cong B \oplus L(A,[BA],[BBA],*) \cr
&\cong B \oplus [BBA] \oplus L(A,[BA],[BBA,A],[BBA,BA],[BBA,BBA,A]),*)}
$$
where the generators now have degrees 2, 1, 1, 0, and 0 respectively.
Continuing in this way we get the following.

\begin{theorem}
Let the degree of $A$ be 2 and the degree of $B$ be 0 with respect
to the polynomial grading (Eq. \ref{eq:degree}). Then
for all $k\ge 0$ we have the following isomorphism,
$$ L(A,B) \cong \Z_k \oplus L(A,\X_k,\Y_k,*)$$
where
\begin{equation}
\label{eq:it}
\eqalign{ \X_0=\emptyset,\quad \X_{k+1} &= \X_k\cup [\Y_k,A], \cr
   \Y_0=\{B\},\quad \Y_{k+1} &= [\Y_k,\X_k]\cup [\Y_k,\Y_k,A] 
= [\Y_k,\X_{k+1}], \cr
   \Z_0=\emptyset,\quad \Z_{k+1} &= \Z_k \cup \Y_k, \cr
}
\end{equation}
and $*$
represents generators which are zero in $\P$, i.e., elements
of the kernel of the homomorphism $L(A,B)\to L_\P(A,B)$.
The generating sets have the following properties:
\begin{enumerate}
\item All elements of $\Y_k$ and $\Z_k$ have degree 0, and all elements of 
$\X_k$ have degree 1.
\item The Lie algebra spanned by $\Z_k$ is abelian.
\item $\X_k = [\Z_k,A]$.
\item All elements of $\Y_k$ and $\Z_k$ have odd order, and all elements
of $\X_k$ have even order.
\item The element of smallest order in $\Y_k$ is 
$(-1)^k [[BA]^kB]$, with order $2k+1$. 
\item The element of largest order in $\Y_k$ is $B_k$, defined recursively
by $B_0=B$, $B_{k+1}=[B_k,[B_k,A]]$. It has order $2^{k+1}-1$.
\item The finite sets $\X_k$ and $\Z_k$ converge to infinite sets
$\Z$ and $\X=[\Z,A]$
in the sense that the sets 
$$ \{X:X\in \X_k, \order(X)\le n\}  $$
are all equal for $k\ge n/2$. We have
$$ L(A,B) \cong \Z \oplus L(A,\X,*)$$
and
\begin{equation}
\label{eq:main}
L_{\P}(A,B) \cong \Z \oplus L(A,\X).
\end{equation}
\item The sizes of the sets $\X_k$ and $\Y_k$ obey the iteration
\begin{equation}
\label{eq:it2}
\eqalign {|\X_{k+1}| &= |\X_k| + |\Y_k| \cr
|\Y_{k+1}| &= |\Y_k|\, |\X_{k+1}| \cr
}
\end{equation}
with initial conditions $|\X_0|=0$, $|\Y_0|=1$. This iteration generates
the sequence of $|\X_k|$ values 
\begin{equation}
\label{eq:a001696}
0, 1, 2, 4, 12, 108, 10476, 108625644,\dots;
\end{equation}
there is a constant $\gamma\approx 1.1555$ 
such that for sufficiently large $k$, $|\X_k| = [\gamma^{2^k}]$.
\end{enumerate}
\end{theorem}
\begin{proof}
The iteration results from successive elimination of elements of degree 0,
each iteration introducing only a finite number of new elements
nonzero in $\P$, which 
have degrees 0 and 1.
The other points then follow easily. 
The final description of $L_\P(A,B)$, Eq. (\ref{eq:main}), follows because
the generators of 
$L(A,\X,*)$ have degree 2 ($A$), 1 ($\X$), or are identically zero
($*$). Therefore $L(A,\X,*)$ 
contains no elements of degree 0, so $L_{\P}(A,\X) = L(A,\X)$.
The sequence of Eq. (\ref{eq:a001696}) is Sloane's sequence 
A001696 \cite{sloane}, which comes from
the same iteration (Eq. \ref{eq:it2}); the reference there to \cite{aho-sloane}
shows how to establish its doubly-exponential growth.
\end{proof}

The rapid growth of the sets $\X_k$ and $\Y_k$ 
means that it is impossible to carry out the
iteration exactly very far. In practice the generating set $\Z$ can be found
up to any order $n$ by dropping any terms of order $>n$ as soon they appear
in $\Y_k$ (i.e., by quotienting all Lie algebras by the ideal consisting 
of all elements of
order $>n$). We then have $\Y_{[(n+1)/2]}=0$ and the iteration terminates.

The results of the six iterations required when $n=12$ are shown
in Table 1. We name the elements of $\Z$ $Z_1$,
$Z_2,\dots$ as they are successively generated by the algorithm. This gives 
a short description of the elements of 
$(L_\P)_n(A,B)$ of order $\le 12$ in terms of 14 elements of degree 0,
14 elements of degree 1, and 1 element of degree 2, which generate
a total of 283 elements of weight $\le 12$ (see Tables 2 and 3).

\begin{table}
\begin{center}
\small
\caption{Elements of degree 0
and weight $\le 11$
(i.e., functions of $q$ only or `modified potentials' of simple
mechanical systems) appearing at iteration $k$ of Eq. (\ref{eq:it}). The new
elements are numbered consecutively $Z_1$, $Z_2,\dots$. The
degree 1 elements $X_n := [Z_n,A]$ also appear.}
\normalsize
\begin{tabular}{|llr|}
\hline
$k$ & $\Y_k$ & order \cr
1&	$Z_1=B$	 & 1\cr
2&	$Z_2=[Z_1,X_1]\quad (=[BBA])$ & 3 \cr
3&	$Z_3=[Z_2,X_1]\quad (=[BBA,BA])$ & 5 \cr
 &	$Z_4=[Z_2,X_2]\quad (=[BBA,[BBA,A]])$ & 7 \cr
4&	$Z_5=[Z_3,X_1]\quad (=[BBA,BA],BA])$ & 7 \cr
 &	$Z_6=[Z_3,X_2]\quad (=[[BBA,BA],[BBA,A]])$ & 9 \cr
 &	$Z_7=[Z_3,X_3]$ & 11 \cr
 &      $Z_8=[Z_4,X_1]$ & 9 \cr
 &	$Z_9=[Z_4,X_2]$ & 11 \cr
5&	$Z_{10}=[Z_5,X_1]$ & 9\cr
 &	$Z_{11}=[Z_5,X_2]$ &11 \cr
 &	$Z_{12}=[Z_6,X_1]$ &11 \cr
 &	$Z_{13}=[Z_8,X_1]$ &11 \cr
6&	$Z_{14}=[Z_{10},X_1]$ &11 \cr
\hline
\end{tabular}
\end{center}
\end{table}

\section{Dimensions of the homogeneous components.}
We now turn to the enumeration of $\X_k$ and $\Y_k$ by order.
We introduce the generating functions
$$ \eqalign{x_k(t) &= \sum_{n=1}^\infty \big|\{X\in \X_k: \order(X)=n\}\big| t^n \cr
   \tilde y_k(t) &= \sum_{n=1}^\infty \big|\{Y\in \Y_k: \order(Y)=n\}\big| t^n \cr
   \tilde z_k(t) &= \sum_{n=1}^\infty \big|\{Z\in \Z_k: \order(Z)=n\}\big| t^n \cr
}$$
which from Eq. (\ref{eq:it}) obey
$$\eqalign{
          x_0 &= 0, \cr
   \tilde y_0 &= 0, \cr
   \tilde z_0 &= t, \cr
          x_{k+1} &= x_k + t \tilde y_k, \cr
   \tilde y_{k+1} &= \tilde y_k x_{k+1}, \cr
   \tilde z_{k+1} &= \tilde z_k + \tilde y_k.\cr
}$$
We can eliminate the $t$-dependence of this map by introducing
$y_k= t \tilde y_k$ and $z_k=t \tilde z_k$. Then 
$z_k\equiv x_k$ for all $k$ and the rest of the system is
\begin{equation}
\label{eq:it3}
\eqalign{
   x_0 &= 0, \cr
   y_0 &= t^2, \cr
   x_{k+1}&= x_k + y_k, \cr
   y_{k+1} &= y_k x_{k+1}. \cr
}
\end{equation}
The polynomials $x_k(t)$ converge to a formal power series
$x(t)$. The polynomials $y_k(t)$ converge, again in the sense of formal
power series, to 0. The power series $x(t)$ completely determines the 
dimensions of the homogeneous components of 
$(L_\P)_n(A,B)$ (including its abelian part $\Z$, because $z_k(t) = x_k(t)/t$).
We find
\begin{equation}
\label{eq:x}
\eqalign{x(t) = &t^2 + t^4 + t^6 + 2t^8 + 3t^{10} + 6t^{12} 
+12 t^{14} + 24 t^{16}
+ 50 t^{18} + 107 t^{20} + 232 t^{22} + \cr
& 508 t^{24} + 1124 t^{26}+
2513 t^{28}+5665t^{30}+12858t^{32}+29356t^{34}+67371t^{36}+\dots}.
\end{equation}
(For example, the $1+1+1+2+3+6=14$ generators of weight $\le 12$
are given in Table 1.)
Amazingly, this power series has appeared before (apparently
as a curiosity) from the 
same iteration (Eq. \ref{eq:it3}), and it appears as Sloane's sequence A045761
\cite{sloane}. 

The classical formula of Witt, Eq. (\ref{eq:witt}), can be extended
to free Lie algebras with more general generating sets
\cite{ka-ki,mu-ow}.
For any set $\A$ with generating function 
$a(t)=\sum_{n>0}|\{A\in\A:\order(A)=n\}|t^n$, the dimensions
$c_n =\dim L_n(\A)$ of the 
homogeneous components of the
graded Lie algebra $L(\A)=\bigoplus_{n>0}L_n(\A)$ are given by
\begin{equation}
\label{eq:cns}
c_n = \sum_{d|n}\frac{1}{d}\mu(d) b_{n/d},
\end{equation}
where
$$ -\log(1-a(t)) = \sum_{n>0} b_n t^n.$$
In Maple, one can compute the dimensions easily by
{\tt c=EULERi(INVERT(a))}
(these functions are available in \cite{sloane}),
where {\tt a} and {\tt c} are the sequences of coefficients of $a(t)$ and
$c(t)$, respectively.

\begin{table}
\small
\begin{center}
\caption{Dimensions of Lie algebras graded by order.
Column 2: Of the free Lie algebra with two generators.
Column 3: Of the Lie algebra of classical mechanics, $L_\P(A,B)$
where $A$ (`kinetic energy') has degree 2 in $p$ and $B$ (`potential
energy') has degree 0 in $p$, i.e. is a function of $q$ only.
Column 4:
Number of modified potentials of order $n$ in $L_\P(A,B)$.
Column 5: Upper bound for maximum number of linearly independent Poisson 
brackets of order $n$ when $M=\R^n$ with the Euclidean metric, i.e.
$A=p^{\rm T} p$. 
Column 6: As Column 5, but $V(q)$ is a cubic
polynomial.}
\begin{tabular}{|rrrrrr|}
\hline
$n$ & $\dim L_n(A,B)$&$\dim(L_\P)_n(A,B)$&$[t^{n+1}]x(t)$& Euclidean & Cubic \cr
\hline
1&      2     &           2 & 1&   2 & 2\cr
2&      1     &           1 & &   1 & 1\cr
3&      2     &           2 & 1&   2 & 2\cr
4&      3     &           2 & &   2 & 2\cr   
5&      6     &           4 & 1&   4 & 3\cr
6&      9     &           5 & &   5 & 3\cr
7&      18    &          10 & 2&   10& 6\cr
8&      30    &          14 & &   14& 6 \cr
9&      56    &          25 & 3&   25&10\cr
10&     99    &          39 & &   39&12 \cr
11&     186   &          69 & 6&   69&19\cr
12&     335   &         110 & &   110&22\cr
13&     630   &         194 & 12&   193&\cr
14&     1161  &         321 & &   320&\cr
15&     2182  &         557 & 24&   555&\cr
16&     4080  &         941 & &   938&\cr
17&     7710  &        1638 & 50&   1631&\cr
18&     14532 &        2798 & &   2787&\cr
19&     27594 &        4878 & 107&   4857&\cr
20&     52377 &        8412 & &  8376&\cr
21&     99858 &       14692 & 232&  14624&\cr
22&    190557 &       25519 & &  25399&\cr
23&    364722 &       44683 & 508&  44460&\cr
24&    698870 &       77993 & &  77594&\cr
25&   1342176 &      136928 & 1124&136191&\cr
26&   2580795 &      240013 & &      238684&\cr
27&   4971008 &      422360 & 2513&419916  &\cr
28&   9586395 &      742801 & &      738375&\cr
29&  18512790 &     1310121 & 5665&130199  &\cr
30&  35790267 &     2310451 & &       2295702&\cr
31&  69273666 &     4083436 & 12858&4056416  &\cr
32& 134215680 &     7218252 & &       7169109&\cr
33& 260300986 &    12781038 & 29356&12691109  &\cr
34& 505286415 &    22638741 & &       22474996&\cr
35& 981706806 &    40152860 & 67371&39853452  &\cr
36&1908866960 &    71247291 & &       70701714&\cr
37&3714566310 &   126559227 & 155345& 125562178 &\cr
38&7233615333 &   224917313 & & 223099566 &\cr
39&14096302710 &  400080000 & 359733& 396759314 &\cr
40&27487764474 &  711997958 & & 705941791 &\cr
\hline
\end{tabular}
\end{center}
\end{table}

We apply Eq. (\ref{eq:cns}) to $L_\P(A,B)\cong \Z\oplus L(A,\X)$. The
generating function for the grading by order of
$\{A\}\cup\X$ is $t+x(t)$. This gives the dimensions
listed in Table 2 for $1\le n \le 40$. A dramatic reduction in the 
dimensions compared to those of the free Lie algebra of rank 2
is evident.

More generally still, Kang and Kim \cite{ka-ki} consider the grading
of a free Lie algebra by an abelian semigroup $S$ which satisfies the
finiteness condition that any $s\in S$ is a sum of other elements of
$S$ in only finitely many ways. Then we have 
\begin{equation}
\label{eq:kk}
\dim L_s(\A) = \sum_{d|s} \frac{1}{d}\mu(d) b_{s/d}
\end{equation}
where
$$ -\log(1-a(t)) = \sum_{s\in S} b_s t^s$$
and $d|s$ means that there exists $\tau\in S$ such that $d\tau=s$,
in which case we write $s/d=\tau$. 

We can use this to calculate the dimensions of $L_{\P}(A,B)$ with 
respect to the bigrading by order and degree.  We first simplify
the grading by degree, Eq. (\ref{eq:degree}), by introducing
$\mathop{\rm degree}'(x) := \mathop{\rm degree}(x) - 1$.
Then (as long as no elements of degree 0 enter, which now holds), the 
semigroup of the grading
by degree$'$ is isomorphic to the nonnegative integers under
addition. Including the grading by order gives
$S={\mathbb Z}^{>0}\times {\mathbb Z}^{\ge 0}$. Note that the finiteness
condition holds for $S$ since it holds for ${\mathbb Z}^{>0}$.
Since $\order(A)=2$, $\degree'(A)=1$, and
$\degree'(X)=0$ for all $X\in\X$, the generating function
of $\{A\}\cup\X$ is $u t + x(t)$ and 
we apply Eq. (\ref{eq:kk}) with
$$b_{t,u} = -[t^n u^m] \log(1-u t - x(t)).$$
This gives the dimensions for $n,m\le 16$ as shown in Table 3.

\begin{table}
\small
\caption{
Dimensions of $L_\P(A,B)$, graded by degree $m$ and by order $n$.}
\begin{tabular*}{\textwidth}%
{@{\extracolsep{0pt minus 5pt}}|r|r|rrrrrrrrrrrrrrrr|}
\hline
&$m$&0&1&2&3&4&5&6&7&8&9&10&11&12&13&14&15 \cr
$n$ & total &  & & & & & & & & & & & & & & &\cr
\hline
1 & 2 & 1& 0 & 1             &&&&&&&&&&&&& \cr
2 & 1 & 0& 1                &&&&&&&&&&&&&& \cr
3 & 2 &  1& 0 &1             &&&&&&&&&&&&& \cr
4 & 2 & 0& 1 &0 &1            &&&&&&&&&&&& \cr
5 & 4 & 1& 0 &2 &0& 1          &&&&&&&&&&& \cr
6 & 5 & 0& 2 &0 &2& 0& 1        &&&&&&&&&& \cr
7 & 10 & 2& 0 &4 &0& 3& 0& 1     &&&&&&&&& \cr
8 & 14 & 0& 4 &0 &6& 0& 3& 0& 1   &&&&&&&& \cr
9 & 25 & 3& 0 &9 &0& 8& 0& 4& 0& 1 &&&&&&& \cr
10 & 39 & 0& 9 &0&14& 0&11& 0& 4&0&1&&&&&& \cr
11 & 69 & 6& 0&20& 0&23& 0&14& 0& 5& 0& 1 &&&&& \cr
12 & 110 & 0&18& 0&37& 0&32& 0&17& 0& 5& 0& 1 &&&& \cr
13 & 194 &12& 0&46& 0&62& 0&46& 0&21& 0& 6& 0& 1   &&& \cr
14 & 321 & 0&42& 0&90& 0&97& 0&60& 0&25& 0& 6& 0& 1 &&  \cr
15 & 557 &24&0&107&0&165&0&144& 0&80& 0&29& 0& 7& 0& 1 & \cr
16 & 941 &0 &90&0& 229&0 &274&0&206&0&100& 0&34& 0& 7& 0& 1 \cr
\hline
\end{tabular*}
\end{table}

\section{Asymptotics of the dimensions and calculation of the entropy.}

From Eq. (\ref{eq:cns}), the asymptotic growth of the dimensions $c_n$ 
is determined by
the analytic structure---the location and type of the singularities---%
of $-\log(1-a(t))$. These correspond to zeros and singularities of
$1-a(t)$. In particular, if $1-a(t)$ has a simple zero at $t=\alpha$
and no other zero with $|t|\le\alpha$, then 
\begin{equation}
\label{eq:cns2}
 c_n \sim \frac{1}{n}\left(\frac{1}{\alpha}\right)^n
\end{equation}
and the Lie algebra has entropy $1/\alpha$.

The generating function of $\{A\}\cup \X$ is $t+x(t)$. We therefore need
to study the analytic structure of the function $1-(t+x(t))$.
We therefore study the map of Eq. (\ref{eq:it3}) considered as a map 
$$\phi:\C^2\to\C^2,\quad (x,y) \mapsto (x+y, y(x+y))$$
with initial conditions $x=0$, $y=t^2$. If the iterates of the map converge
to $(x^*,0)$ say, then $x(t) = x^*$. Curiously, the map preserves
the area $\frac{1}{y}dx\wedge dy$, although this plays no role in the analysis.

The map $\phi$ has a line of degenerate fixed points $(x,0)$ with
eigenvalues $x$ and 1. The fixed points with $|x|>1$ are unstable and
one can show that the fixed points with $|x|<1$
are stable.
The map `remembers' its initial condition, and the function $x(t)$
is the $x$-coordinate of the fixed point reached from initial condition
$(0,t^2)$. 

We can see immediately that (i) for $t$ real and positive, 
$x(t)$ is strictly increasing; and (ii) if the map converges then
$|x(t)|\le 1$. For $t$ real and positive, the sequence
$\{y_k\}$ is increasing, and if there is a $k$ such that
$y_k>1$, then $x_k\to\infty$. Therefore we define
$$\beta=\inf\{t\in\R^+\colon\  x_k(t)\to\infty\}.$$
Because
$$ |x_{k+1}| \le |x_k| + |z_k|,\quad |z_{k+1}| = |z_k| |x_{k+1}|,$$
the map converges in the disk $\{t: |t|<\beta\}$.

We can get a crude bound on $\beta$ immediately, but more detailed
knowledge requires a numerical study of the map $\phi$. Let
$t$ be real, let $I(x,y)=y+x-1+\sqrt{2y}$, and suppose $x>0$, $y>0$, and
$I(x,y)<0$. 
Then 
$$\eqalign{I\circ\phi - I &= 
y(x+y)+x+y-1+\sqrt{2y(x+y)}-(y+x-1+\sqrt{2y}) \cr
&= y(x+y)+\sqrt{2y}(\sqrt{x+y}-1) \cr
&\le y(1-\sqrt{2y}) + \sqrt{2y}\sqrt{1-\sqrt{2y}}-1) \cr
&\le y(1-\sqrt{2y}) + \sqrt{2y}(1-\frac{1}{2}\sqrt{2y})-1) \cr
&= -\sqrt{2}y^{3/2} \cr
&<0\cr
}$$
Therefore, the orbit must stay in the bounded region
$x>0$, $y>0$, $I(x,y)<0$, with $x_k$ increasing and $y_k$ decreasing.
Therefore the orbit converges to some fixed point $(x,0)$.
(Here the curve $x=1-\sqrt{2y}-y$ was chosen because it is a good approximation
of the stable manifold of $(1,0)$.)
Since $I(0,t^2)<0$ for $0<t^2<2-\sqrt{3}$, we have $\beta>\sqrt{2-\sqrt{3}}>
0.51$.
Better approximations of $\beta$ can be obtained as the roots of
$I\circ\phi^k(0,\beta^2)=0$ 
(i.e., by requiring the $k$th iterate to land in the 
trapping region), but these must be calculated numerically.
On the other hand, $x_2=t^2+t^4>1$ if $t>0.79$, so we have the bounds
$0.51<\beta<0.79$.

We have that $dx(t)/dt>0$
on $[0,\beta)$, with 
$x(0)=0$ and $x(\beta)=1$;
and $1-t$ is decreasing. Therefore $1-t-x(t)$ has exactly one zero
in $[0,\beta)$, and it is simple. 
The zero is $\alpha$, the reciprocal of the required entropy of
$L_\P(A,B)$. The numerical value of $\alpha$ can be determined by
solving $1-t-x(t)=0$ numerically.%
\footnote{In {\sc Matlab}, by {\tt function x = f(t); x=0; y=t\^{ }2;
while y>1e-16, x=x+y; y=y*x; end; x = 1-t-x;} and
{\tt alpha = fsolve('f',0.5)}.}
This gives the value of the entropy of $L_\P(A,B)$ as
\begin{equation}
\label{eq:alpha}
 1/\alpha = 1.82542377420108\dots.
\end{equation}

Are there any other solutions to $1-t-x(t)=0$? 
Because the coefficients of $x(t)$ are all nonnegative, there
can be none in the disk $|t|\le\alpha$.
To say more we have to proceed numerically. Firstly, if $|x_k|$ 
and $|y_k|$ get too large then the orbit blows up. Let 
$$D=\{(x,y)\in \C^2\colon |y|>2|x|>2\}.$$
Suppose $(x_k,y_k)\in D$.
Then
$$ |x_{k+1}| \ge | |y_k| - |x_k| | > |x_k| > 1 $$
and 
$$ |y_{k+1}| = |x_{k+1}| |y_k| > 2|x_{k+1}|,$$
i.e., we have $(x_{k+1},y_{k+1})\in D$.
The orbit then stays in $D$ and cannot converge---in fact, 
it must blow up doubly exponentially. The first iterate
$(x_1,y_1)=(t^2,t^4)$ is in $D$ if $|t|>\sqrt{2}$, and
the second iterate $(x_2,y_2)=(t^2+t^4,t^4(t^2+t^4))$ is in $D$ if 
$|t|>1.27202$. 
In practice, if an iterate enters this region one can immediately
stop the calculation and report that the map diverges. 

\begin{figure} \begin{center}
\vbox{ \includegraphics[width=11cm,keepaspectratio,clip]{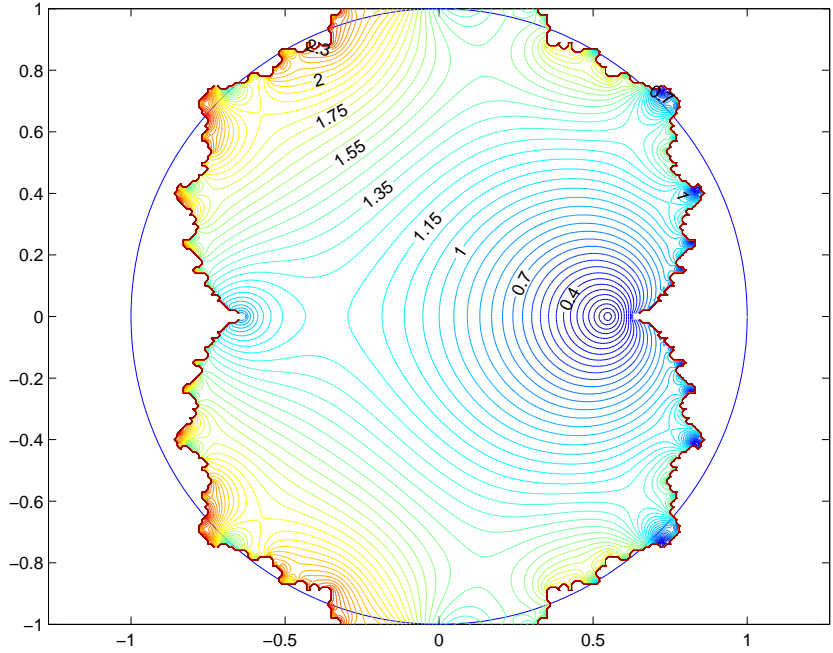} }
\caption{Contour plot of $|1-t-x(t)|$, showing its main zero
at $t=\alpha=1/1.8254\dots$ and other zeros (two sequences
approaching $t=\beta$. The unit circle is also shown.}
\label{fig3.ps}
\end{center} \end{figure}

\begin{figure}\begin{center}
\hbox{\hskip0cm\includegraphics[width=8cm,keepaspectratio,clip]{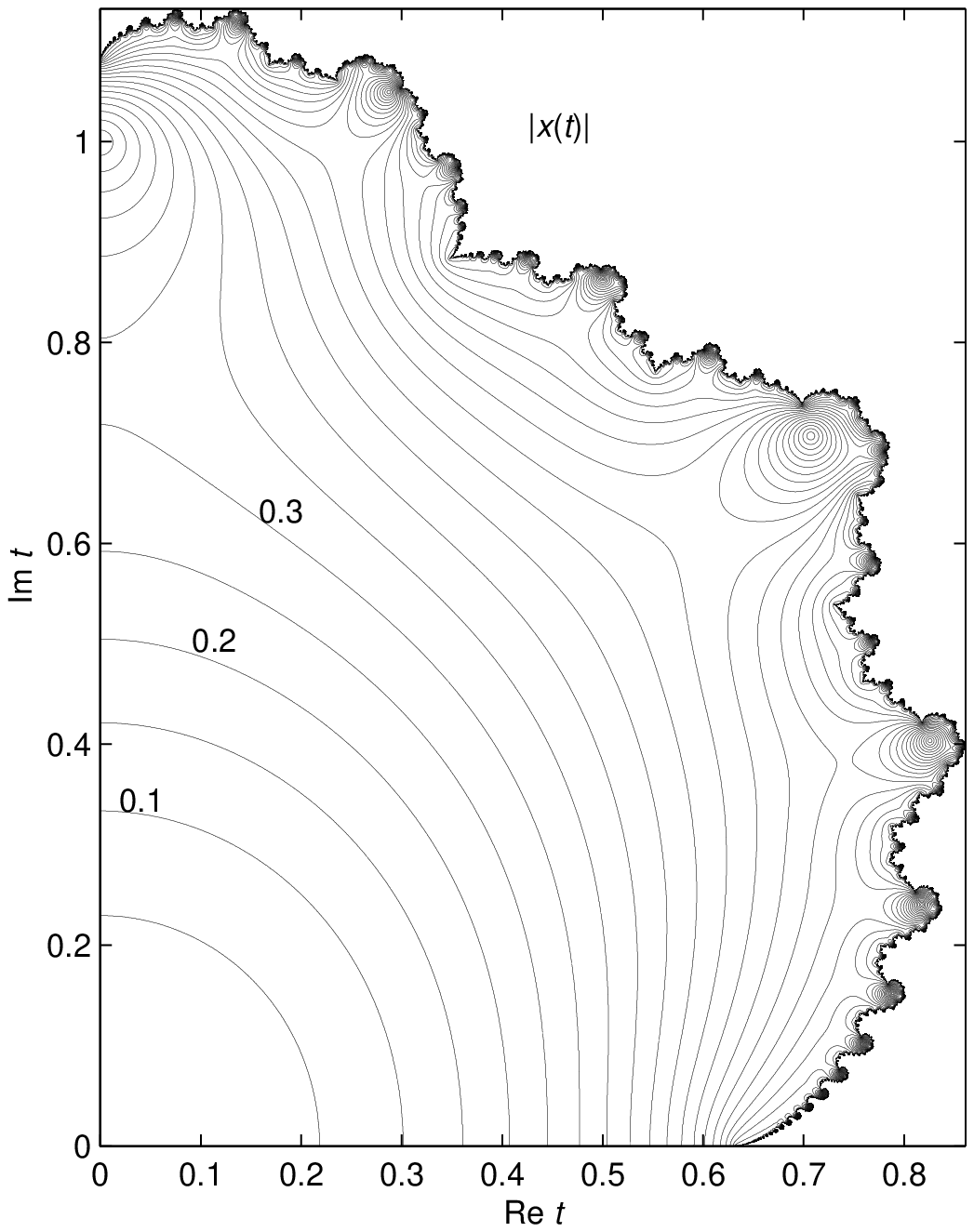}
\includegraphics[width=8cm,keepaspectratio,clip]{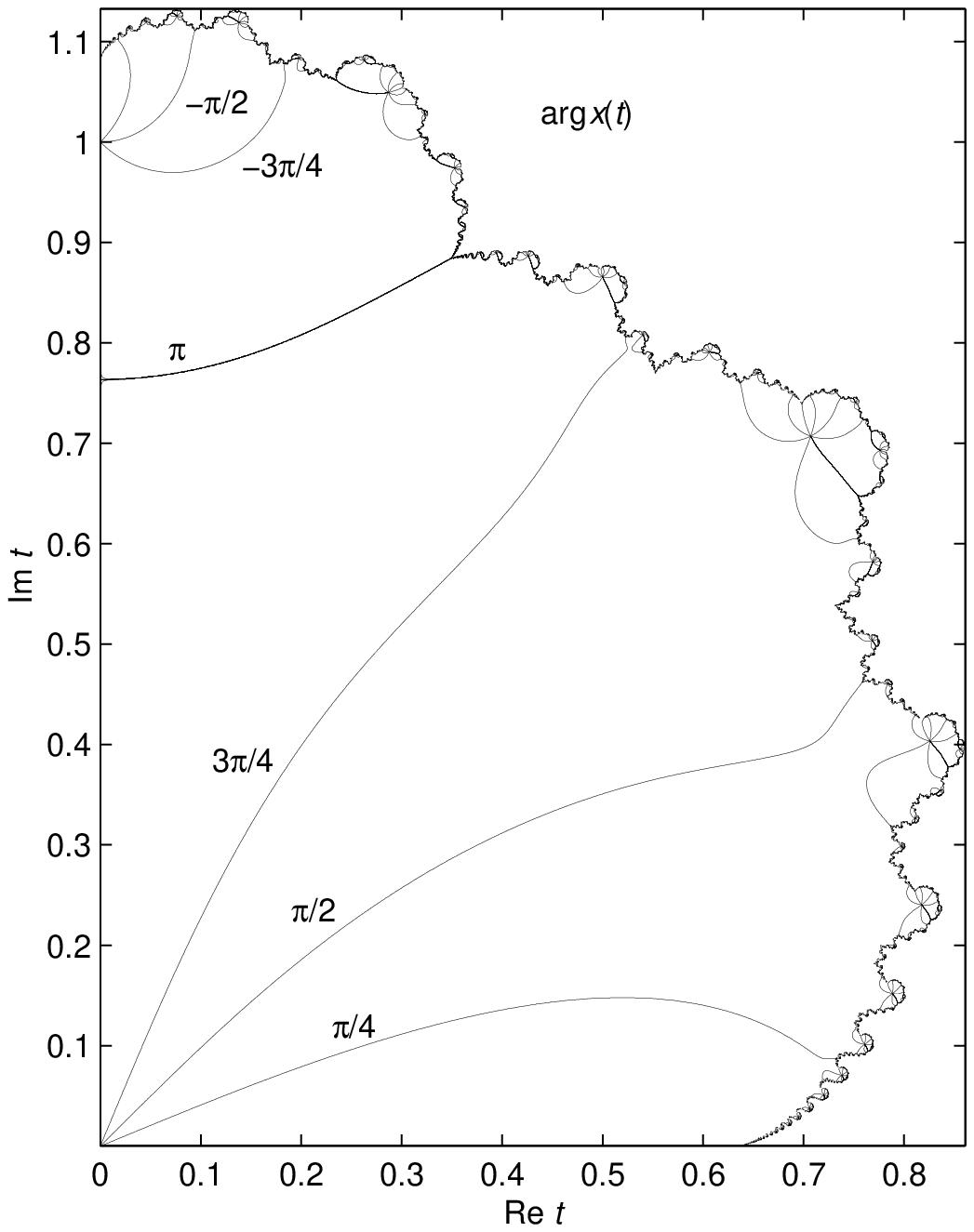}}
\caption{Contour plots of $|x(t)|$ (left, contour interval 0.05) and arg$x(t)$ (right).}
\label{modf.eps}
\end{center}
\end{figure}

Using this criterion we computed the function
$x(t)$ numerically. See  Figures 1 and 2.

We have made the following numerical observations:
\begin{enumerate}
\item The singularity of $x(t)$ closest to the origin is at
\begin{equation}
\label{eq:beta}
t=\beta = 1/1.58207912734\dots.
\end{equation}
\item There are no zeros of $1-t-x(t)$ in the disk $|t|<\beta$.
\item The map converges only in a connected, simply-connected
region with a fractal boundary.
\item The function $x(t)$ is analytic everywhere inside this
region but has a square root singularity everywhere on its boundary.
\item For each point $z$ on the boundary, $x(t)\sim 1 - a (t-z)^{1/2}$
for some constant $a$ depending on $z$, as $t\to z$.
\item There is only one zero of $1-t-x(t)$ in $|t|\le\beta$.
\item The other zeros of 
$1-t-x(t)$ form two infinite sequences $\alpha_n$, $\bar\alpha_n$,
with $Re(\alpha_n)>\beta$ for all $n$ and $\lim_{n\to\infty}\alpha_n=\beta$.
\end{enumerate}
Because of the fractal nature of the boundary, we are unlikely to be 
able to `solve' the map $\phi$ or find $\alpha$ in closed form.
Observation (4) would imply that this boundary forms a natural
boundary for the function $x(t)$.
Observation (5) would imply that the number of modified potentials of
order $n$,
$[t^{n+1}]x(t)$, is $\O(n^{-3/2}\beta^{-n})$.
Observation (6) would imply that the next term in the asymptotic
growth of $c_n=\dim (L_\P)_n(A,B)$ comes from the square root singularity
at $t=\beta$. Indeed, by computing $c_n$ numerically for $n<80$ we
find that 
$$c_n\sim n^{-1} \alpha^{-n} - 
\left\{\matrix{1.51 n^{-1/2} \beta^{-n} & n \hbox{\rm\  odd} \cr
1.61 n^{-3/2} \beta^{-n} & n \hbox{\ even}\cr}\right.$$
and
$$ [t^{n+1}]x(t) \sim 0.9628 n^{-3/2} \beta^{-n} $$
for $n$ even. These are all consistent with the observed singularity
structure of $1-t-x(t)$.

\section{Discussion}
\subsection{Physical interpretation of the generators}

There is a particularly nice interpretation of $L_\P(A,B)\cong \Z\oplus
L(A,\X)$ in the
specific case of simple mechanical systems. In local coordinates,
let $A=T(p)=\frac{1}{2}p^T M(q) p$ be the kinetic energy, where
$M(q)$ is the inverse of the metric (or mass matrix), and 
and $B=V(q)$ be the potential energy. The set $\Z$ consists
of functions of $q$ only, and we think of them as `modified
potentials'. Elements of the span of $\Z$, 
$$\sum_{Z\in \Z}a_Z\tau^{\degree(Z)} Z = a_1 \tau V + a_2 \tau^3 M(V',V')
+ \dots, $$
and their flows, can be evaluated explicitly and used to construct
high-order integrators of the full system $T+V$ (see Eq. (\ref{eq:mps})
for more terms). Now consider the generator $X=[Z,A]\in\X$. It is the
cotangent lift of the gradient flow of the modified potential $Z$; 
we have $X=M(q)(Z(q),p)$ and Hamilton's equations are
$$\eqalign{ \dot q &= M(q)Z'(q) = \div_{M^{-1}(q)} Z =: f(q) \cr
   \dot p &= -f'(q)^{\rm T}p. \cr
}$$
So in a sense the modified potentials and the kinetic energy together
contain a complete description of the Lie algebra.

\subsection{Euclidean mechanical systems}
Recall that on each manifold $M$, each simple mechanical system 
(say with kinetic energy $T$ and potential energy $V$) 
generates a Lie algebra of class $\P$. Therefore there
is a homomorphism $\psi(M,T,V)$ from $L_\P(A,B)$ onto this Lie algebra.
One can ask whether the system $(M,T,V)$ is in general position,
i.e. if the two Lie algebras are actually isomorphic and $\ker\psi(M,T,V)=0$.
This is unlikely, because of the existence of identities such 
as Eq. (\ref{eq:standard4}) in Lie algebras of vector fields. One can therefore
consider larger {\it classes} of systems and ask whether they
are in general position. That is, does the class satisfy any
identities other than those corresponding to the grading
by degree, Eq. (\ref{eq:degree})? We conjecture that for
the class of all simple mechanical systems, it does not.
\begin{conjecture}
\label{con:main}
The only identities satisfied by all simple mechanical systems
are those due to the grading by degree. That is, 
$$ \bigcap_{M,T,V}\ker\psi(M,T,V) = 0. $$
\end{conjecture}

This is best discussed by introducing a smaller class which we shall
see is {\it not} in general position. Namely, let $M=\R^n$ with
the Euclidean metric. Then in coordinates the kinetic energy is
$T(p) = \frac{1}{2}\sum_{i=1}^n p_i^2$. The first few modified potentials
are then
$$\eqalign{ Z_1 &= V \cr
Z_2 &= [BBA] = V'(V') \cr
Z_3 &= [BBA,BA] = 2V''(V',V') \cr
Z_4 &= [BBA,[BBA,A]] = 4V''(V''(V'),V')\cr
Z_5 &= [[BBA,BA],BA] = 2V'''(V',V',V')+4V''(V''(V'),V') \cr
}
$$
where we regard the $k$th derivative of $V$ as a real-valued symmetric linear
function on $k$ vectors. Each modified potential of order $2n-1$ is a linear
combination of the scalar elementary differentials of order $n$ of $V$.
Each such differential can be associated to a free tree with $n$
nodes. 
(See, for example, \cite{ha-lu-wa} for a discussion of elementary
differentials and trees.)
The number of such trees for $n\ge 1$ is (Sloane's A000055, 
\cite{sloane})
1, 1, 1, 2, 3, 6, 11, 23, 47, 106, 235, \dots. This should be compared
with the number of modified potentials in Eq. (\ref{eq:x}), namely
1, 1, 1, 2, 3, 6, 12, 24, 50, 107, 232, \dots.
There are three interesting consequences:
\begin{itemize}
\item[(i)]
For $n\le 6$, the sequences are the same. In fact, one can check
that in the modified potentials of orders $2n-1\le 11$, all 
trees appear, in invertible linear combinations, so these modified
potentials are in general independent.
\item[(ii)] For $n=7,8,9$, there are more modified potentials than free trees.
In particular, only 11 of the 12 modified potentials of order 13
can be linearly independent. This proves that the class of
{\it Euclidean} mechanical systems is not in general position.
\item[(iii)] For $n\ge 10$, there are fewer modified potentials than free trees.
In fact, the former have entropy $1/\beta=1.582\dots$ while the latter
(since the free trees have entropy given by Otter's constant,
2.955\dots)
have entropy $\sqrt{2.955\dots}=1.719\dots$. 
Thus, for large $n$, only certain combinations
of the trees appear in $\Z$.
\end{itemize}

So far we have only considered
the modified potentials $\Z$ 
themselves. If these are independent, then $\X=[\Z,A]$ is independent too.
However, there is still a possibility for extra identities to hold
in the Lie algebra generated by $A$ and $\X$.
A term of order $n$ and degree $m$ is a sum of 
elementary differentials of $V$ and $p$, corresponding to
trees with $(n+m+1)/2$ nodes, of which $m$ leaves are labelled $p$
and the remaining nodes are labelled $V$. In this case we find
that for $(n+m+1)/2\le 7$ there are always sufficient labelled
free trees to prevent forced dependencies among the Lie brackets.
For example, of the 11 free trees with 7 nodes, there are
12, 20, 24, 18, 9, 3, and 1 trees in which $m=0$, 1, 2, 3, 4, 5, and 6 leaves
are coloured $p$, respectively. The dimensions of the
corresponding homogeneous subspaces of $L_\P(A,B)$ with 
$(n+m+1)/2=7$ are (from Table 3) 12, 18, 20, 14, 8, 3, and 1, respectively.
Thus, only in the case $m=0$, corresponding to the modified potentials
themselves, is a dependency forced in this way.

The algorithm for $L_\P(A,B)$, Eqs. (\ref{eq:it}) and (\ref{eq:it3}),
can be modified to take into account the
dependencies amongst the Lie brackets in the Euclidean case.
To get an upper bound on the dimensions and entropy of the Lie algebra
in this case, we assume that the dependency appears only when
forced. Let $c_n$ be the number of free trees with $n$ nodes. 
At iteration $k$, we already have $z_{k,2n}:= [t^{2n}]\tilde z_k$
elements of order $2n-1$ in 
$\Z_k$, and $y_{k,2n}:=[t^{2n}]\tilde y_k$ elements
of order $2n-1$ have just been created in $\Y_k$. If $z_{k,2n}+y_{k,2n}>c_n$,
we replace $\Y_k$
by a smaller set, of $c_n-z_{k,2n}-y_{k,2n}$ elements, which together
with the order $2n-1$ elements of $\Z_k$, forms a basis of the $c_n$
elementary differentials. In terms of the generating functions,
we add the final step to the iteration of Eq. (\ref{eq:it3}):
\begin{equation}
\label{eq:new}
y_{k+1} \leftarrow \sum_{n\ge 1} \min(y_{k+1,2n}, c_n-z_{k+1,2n}) t^{2n}
\end{equation}
Let the resulting limiting formal series be $x_E(t)$, $y_E(t)$, and
$z_E(t)$.
The generating function for $x_E(t)$ is then computed to be
\begin{equation}
\label{eq:xe}
\eqalign{x_E(t) = &t^2 + t^4 + t^6 + 2t^8 + 3t^{10} + 6t^{12} 
+11 t^{14} + 23 t^{16}
+ 47 t^{18} +
102 t^{20} + 221 t^{22} + \cr
& 484 t^{24} + 1069 t^{26}+ 2386 t^{28}+ 
5364t^{30}+12143^{32}+27645t^{34}+63259t^{36}+\dots}
\end{equation}
which should be compared with Eq. (\ref{eq:x}). At order
14, 16, and 18 the dimensions are limited by the number of elementary
differentials, but for $n>9$, $[t^{2n}]x_E(t)<[t^{2n}]x(t)<c_n$.
Because the new map on generating functions, Eq. (\ref{eq:new}),
is not analytic, it is harder to determine the location of its
smallest singularity. We found the smallest root of successive
polynomial truncations of $1-t-x_E(t)$ and extrapolated these results
to obtain 
\begin{equation}
\label{eq:alphaE}
1/\alpha_E = 1.8250339\dots,\quad 1/\beta_E = 1.574\dots.
\end{equation}
These are upper bounds for the entropy of the class of Euclidean 
mechanical systems and their modified potentials, respectively.

(Murua \cite{murua} has also considered this case, in the context
of order conditions for Hamiltonians of the form $\frac{1}{2}\sum p_i^2+
V(q)$. He
finds a unique independent tree of a certain type for each order
condition, and enumerates these up to order 6. It would be
interesting to compare the two approaches at higher order.)

The situation is quite different for non-Euclidean, i.e. general,
mechanical systems.
Repeating the above calculation for a general kinetic energy
$T(p)=\frac{1}{2}p^T M(q)p$, we get the following modified
potentials. The associated trees will be explained below.
\begin{equation}
\label{eq:mps}
\eqalign{ Z_1 &= V = \global\def\path{#1}\input{tree11.inp}\cr
Z_2 &= [BBA] = M(V',V') = \global\def\path{#1}\input{tree31.inp}\cr
Z_3 &= [BBA,BA] = 2 M(V',V''(M(V'))) + M'(V',V',M(V')) 
	= 2\global\def\path{#1}\input{tree51.inp} + \global\def\path{#1}\input{tree52.inp} \cr
Z_4 &= [BBA,[BBA,A]]  \cr
    & = 4 V'(M(V''(M(V''(M(V')))))+3V'(M(V''(M(M'(V',V'))))) +\cr
    &	\qquad M(M'(V',V'),M'(V',V')) \cr
    & = 4\,\global\def\path{#1}\input{tree71.inp} + 3\global\def\path{#1}\input{tree72.inp} + \global\def\path{#1}\input{tree73.inp} \cr
Z_5 &= [[BBA,BA],BA] \cr
&=4V'(M(V''(M(V''(M(V'))))))+2V'''(M(V'),M(V'),M(V'))+ \cr
    & \qquad 6 M'(V',M(V'),V''(M(V')))+ M'(V',V',M(V''(M(V')))) + \cr
    & \qquad M'(V',M(V'),M'(V',V')) + M''(M(V'),M(V'),V',V') \cr
    & = 4\,\global\def\path{#1}\input{tree71.inp} + 2\global\def\path{#1}\input{tree74.inp} + 6\global\def\path{#1}\input{tree75.inp}
+ \global\def\path{#1}\input{tree72.inp} + \global\def\path{#1}\input{tree76.inp}
+ \global\def\path{#1}\input{tree78.inp} \cr
}
\end{equation}
In this case
each modified potential of order $2n-1$ 
is a scalar elementary differential of $V$
{\it and} $M$. These correspond to bicoloured 
free trees with $2n-1$ nodes, of which
$n$ nodes are labelled $V$ (shown as solid circles above) and $n-1$ nodes are 
labelled $M$ (shown as open circles above); the latter must have
at least 2 branches since a derivative of $M$ has at least 2 indices.
Of the 1, 1, 3, 11, 47, and 235 free trees of order 1, 3, 5, 7, 9, and 11
respectively, exactly 1, 1, 2, 8, 34, and 175 of them
can be coloured (labelled) in this way. The calculation above
shows that of these colourable trees, precisely one colouring of
1, 1, 2, and 7 of these colourable trees occur
in the modified potentials of orders $\le 7$. (The other colourings
of these trees do not occur, because of the way in which the trees
at each order are built from the trees of lower order. The
colourable 7-node tree \global\def\path{#1}\input{tree77.inp} also does not occur.) 
It is clear that there is enormously much more freedom in this
case than in the (`Euclidean', $T(p)=\frac{1}{2}\sum p_i^2$) case considered
previously. Therefore, we believe that all the modified potentials 
are independent in this case.
This supports Conjecture \ref{con:main}.

\subsection{Other polynomially graded Lie algebra}
We close with a list of some other Lie algebras of class $\P$. In each
case one can consider the case of two generators $A$ and $B$ of
degrees 2 and 0 and the induced homomorphism from $L_\P(A,B)$.
\begin{enumerate}
\item The case of classical mechanics. The objects are
real functions on a cotangent bundle, homogeneous polynomial in $p$. 
This can be specialized to 
the following cases.
\begin{enumerate}
\item $Q$ any Riemannian manifold, any potential energy,
$\degree(X)$ is the total degree of $X$ in $p$. 
Entropy is $\le 1.8254\dots$, Eq. (\ref{eq:alpha}), with
Conjecture \ref{con:main} implying equality.
\item $Q=\R^n$ with the Euclidean metric. Entropy is $\le 1.8250\dots$,
Eq. (\ref{eq:alphaE}). It is remarkable that these two Lie algebras,
not previously distinguished from each other in the literature, differ starting
at order 13, and have slightly different entropy.
\item $Q=\R^n$, functions
polynomial in $p$ {\it and} $q$. We can then introduce
a bigrading by degree in $p$ and by degree in $q$. To get a new Lie algebra,
one of the generators has to be degree 0 in each grading, which forces
$Q$ Euclidean, $A=\frac{1}{2}\sum p_i^2$, $B=V(q)$ polynomial. For example,
we have computed the dimensions of the Lie algebra generated
by cubic potentials for small $n$ in Table 2---they are remarkably small. 
See \cite{sofroniou} for an analysis of this case in terms of special types
of trees.
\end{enumerate}
\item Homogeneous polynomial vector fields on $\R^m$ graded by total
degree in $x_1,\dots,x_k$ for some $1\le k \le m$. In the case $k=m$, 
the vector fields in $\X$ associated with $L(A,B)$ ($\degree(A)=2$, 
$\degree(B)=0$) are associated with free trees in which each
node has degree at most 2 (since only the first two derivatives
of $A$ are nonzero). Their numbers are 1, 1, 1, 2, 3, 6, 11, 23 (so
far the same as for the free trees), then 46, 98, 207, 451, \dots.
(Sloane's A001190 \cite{sloane}), which gives an upper bound
for the number of independent elements of $\Z$ of each odd order.
These grow more slowly than the free trees, and even more slowly
than $\Z$, with entropy 1.5758, compared to 1.5821 (Eq. (\ref{eq:beta}))
for $\Z$. Perhaps in this case the trees $\mathfrak T$  generate the Lie algebra
as ${\mathfrak T} \oplus L(A,[{\mathfrak T},A])$?
\item As the previous item, but multigrading by total degree
in different subsets of the variables.
\item Homogeneous polynomial vector fields with the variables partitioned
$(x,y)$ with $x\in\R^k$, $y\in\R^m$, and the vector fields
of the form $p \frac{\partial}{\partial x} + q\frac{\partial}{\partial y}$
with either $\degree_y(q) \le \degree_y(p)+1$, 
or $p\equiv 0$ and $\degree_y(q)=0$. Simple mechanical systems 
form examples of this class. So do high-order ODEs of the form
$y^{(n)} = f(y,\dots,y^{(n-2)})$ when re-written as first-order systems
$$\eqalign{ \dot x_i &= x_{i+1}, \qquad i = 0,\dots,n-2, \cr
   \dot x_{n-1} &= f(x_1,\dots,x_{n-2}), \cr
}$$
with $x_i=y^{(i)}$, $k=n-1$, and $m=1$.
\item Consider the Schr\"odinger equation 
$$ i\dot\psi = \nabla^2 \psi + V(x)\psi,$$
where $\nabla^2$ is the Euclidean Laplacian.
The two operators $\nabla^2$ and $V(x)$ generate a Lie algebra
of class $\P$, where the grading is by degree of the differential
operators. For example,
$$ [\nabla^2,V]\psi = \nabla\cdot(V\psi) + V \nabla\cdot\psi$$
is of degree 1,
$$ [V,V,\nabla^2]\psi = (\nabla\cdot(V^2))\psi$$
is of degree 0, and
$$ [V,V,V,\nabla^2]\psi \equiv 0.$$
\end{enumerate}

\bigskip
\noindent{\bf Acknowledgements.} 
We would like to thank Mark Sofroniou and Reinout
Quispel for pointing out the connection to 
free trees. RM would like to thank Ernst Hairer and Gerhard
Wanner for their hospitality at the University of Geneva where this
paper was written.

\end{document}